\documentclass{bvm} % Do NOT change this line

\begin{document}
% Do not change anything in the preamble (anything above \begin{document}) except for the specification of the bibliography file, any additional changes will be lost

% You can create your own commands using \newcommand. These must be placed after(!) \begin{document} and should contain your own name to avoid multiple definitions between different papers
\newcommand{\bvmyear}{2025}

% use the \selectlanguage command to select the language in which your proceedings are written
%\selectlanguage{ngerman} % German
\selectlanguage{english} % English

% The title of your paper
\title{Automation Bias in AI-Assisted Medical Decision-Making under Time Pressure in Computational Pathology}

% Alternative ideas
% \title{Do Automation Bias and Time Constrain AI-Aided Decision-Making in Computational Pathology}

% If you write a short paper/abstract, the title must start with "Abstract:".
% \title{Abstract: Bildverarbeitung für die Medizin \bvmyear}

% Optional specification of a subtitle
% \subtitle{Guidelines for the Creation of the Print-ready Contributions}

% titlerunning appears in the header of every second page
% LaTeX generates this automatically from your paper title
% However, if it is too long, the message "Title Suppressed Due to Excessive Length" appears instead.
% In this case, specify an abbreviated form of the title here
\titlerunning{Automation Bias in AI-Ass. Medical Decision Making}

% Please indicate all authors involved
% To allow us to correctly identify the last name of each author, indicate it using the \lname{} command.
% If more than one institute is involved, list the number of the institute(s) (see below) with \inst{} after the respective author. If only one institute is involved, omit this.
% Separate all authors with a comma
\author{
	Emely \lname{Rosbach} \inst{1},
        Jonathan \lname{Ganz} \inst{1}, 
        Jonas \lname{Ammeling} \inst{1},
        Andreas \lname{Riener} \inst{1}, and
	Marc \lname{Aubreville} \inst{2}
}

\authorrunning{Rosbach et al.}

% Specify the institutes involved
% In case of participation of more than one institute, each institute shall be preceded by an ascending number with \inst{}.
% If only one institute is involved, omit the corresponding number.
% Separate individual institutes with \\

\institute{
\inst{1} Technische Hochschule Ingolstadt, Ingolstadt, Germany\\
\inst{2} Flensburg University of Applied Sciences, Flensburg, Germany}
\email{emely.rosbach@thi.de}

\maketitle

% Abstract of your paper, only for long papers
% Do NOT use \begin{abstract} ... \end{abstract} for short articles
\begin{abstract}
 Artificial intelligence (AI)-based clinical decision support systems (CDSS) promise to enhance diagnostic accuracy and efficiency in computational pathology. However, human-AI collaboration might introduce automation bias, where users uncritically follow automated cues. This bias may worsen when time pressure strains practitioners’ cognitive resources. We quantified automation bias by measuring the adoption of negative system consultations and examined the role of time pressure in a web-based experiment, where trained pathology experts (n=28) estimated tumor cell percentages. Our results indicate that while AI integration led to a statistically significant increase in overall performance, it also resulted in a 7\% automation bias rate, where initially correct evaluations were overturned by erroneous AI advice. Conversely, time pressure did not exacerbate automation bias occurrence, but appeared to increase its severity, evidenced by heightened reliance on the system's negative consultations and subsequent performance decline. These findings highlight potential risks of AI use in healthcare.
\end{abstract}

\section{Problem Statement and Related Work}
Driven by the success of deep learning algorithms in medical imaging, computational pathology seeks to augment practitioner capabilities in areas traditionally challenging for humans like quantitative image analysis tasks e.g., manual biomarker scoring. Given the safety-critical nature of healthcare and the complexity of legal liability for machine misjudgments, clinical decision support systems (CDSS) allow the full benefits of artificial intelligence (AI) integration, including improved diagnostic accuracy and increased efficiency, while keeping the responsibility for the final diagnosis with the medical expert. However, the necessity for practitioner oversight harbors the risk of introducing a new set of challenges, as the mere presence of AI advice could trigger or amplify cognitive biases, which are systematic patterns of deviation from rationality in judgment, such as automation bias (AB). AB refers to the tendency to treat automated cues as infallible, following them unquestioningly instead of vigilant information seeking. This leads to errors when issues go unnoticed because the system fails to detect them (omission errors) or when incorrect automation output is uncritically adopted (commission errors)~\cite{Goddard2012}. Environmental factors like time pressure, ubiquitously present in routine pathology, can place strain on cognitive resources, resulting in heuristic-based usage of decision support systems (DSS) or even automation overdependence~\cite{Goddard2012}. In essence, time stress may amplify both the frequency and magnitude of AB. While most studies on CDSS highlight the overall performance gains from AI integration, there have been few studies on the effect size of AB in medical decision-making~\cite{Goddard2012}, with none in the field of pathology. While existing research typically evaluates overall acceptance of false AI advice, following Goddard et al.~\cite{Goddard2012} we will measure AB through commission errors, where a previously correct independent evaluation is overturned by incorrect AI guidance (negative consultation), allowing us to isolate AB from other cognitive biases for more precise quantification. Research on the impact of time pressure on AB presents conflicting results, suggesting its effects may be context-dependent: some work demonstrates increased automation dependence under time constraints~\cite{Rice2008}, while others report reduced system reliance in time critical situations~\cite{Rieger2022}.

\section{Methods}
In this paper, for the first time, we quantify AB in human-AI collaboration within computational pathology and examine the additional influence of time pressure through a web-based experiment conducted with non-paid expert users (pathologists, pathology residents, and non-physician pathology staff) independently verified by us and recruited from our professional network, based on prior research collaborations.
\subsection{Study Task and Materials}
\textit{Study task.} Our study task focused on estimating tumor cell percentage (TCP) on hematoxylin and eosin (H\&E)-stained tissue slides, defined as the ratio of neoplastic cells to the total cell count. Typically represented by a single percentage value or specified range, TCP is estimated briefly through visual assessment in clinical practice and is thus suitable to be performed under time constraints. Due to its innate complexity, stemming from the variability of clinical specimen, lack of standardized protocols and absence of formal training, TCP is widely acknowledged as being prone to substantial inter-observer variability \cite{Smits2014}, rendering it susceptible to cognitive bias manifestation. TCP estimation is commonly performed in routine pathology, as certain molecular tests require a specific level of tumor DNA to ensure robustness and proper interpretation of assay results \cite{Smits2014}. Should samples falling below the threshold for a given molecular test be estimated to be above it, unwarranted confidence in the assay result may be fostered, leading to misguided treatment plans and compromised patient care \cite{Smits2014}.

\textit{Materials.} For the TCP estimation, we provided participants with a selection of 23 tissue patches, presenting a broad spectrum of tumor cellularity and tissue types at different magnifications. Of these images, three were allocated for a training session, while the remaining 20 were designated for the main experiment. The study material\footnote[1]{The final dataset, including a table with image patch details, participant demographics, and data analysis source code are available at: \url{https://anonymous.4open.science/r/AB-TP-6440/}} was sourced from three openly available datasets, each featuring image patches and dense annotations of various cell types, including tumor cells: the BreCaHad dataset~\cite{BreCaHad}, the dataset from Frei et al.'s publication on tumor cell fraction scoring \cite{Frei2023}, and the BreastPathQ dataset \cite{Martel2019}. A standard object detection approach based on the FCOS \cite{Tian2019fcos} architecture was utilized to detect tumor- and other cells, from which the TCP for each slide was calculated. This approach was trained on the BreCaHad dataset, reserving five patches for use in the experiment. Using an actual AI prevented bias from fabricated estimations, allowing us to observe participant reactions to realistic inaccuracies. When the AI model was applied to the study slides, about half showed highly accurate detections, while the remaining contained samples with both high positive- and negative error rates, indicating a realistic non-robustness caused by a covariant data shift between the datasets.
\begin{figure}[h]
    \centering\includegraphics[width=0.70\linewidth]{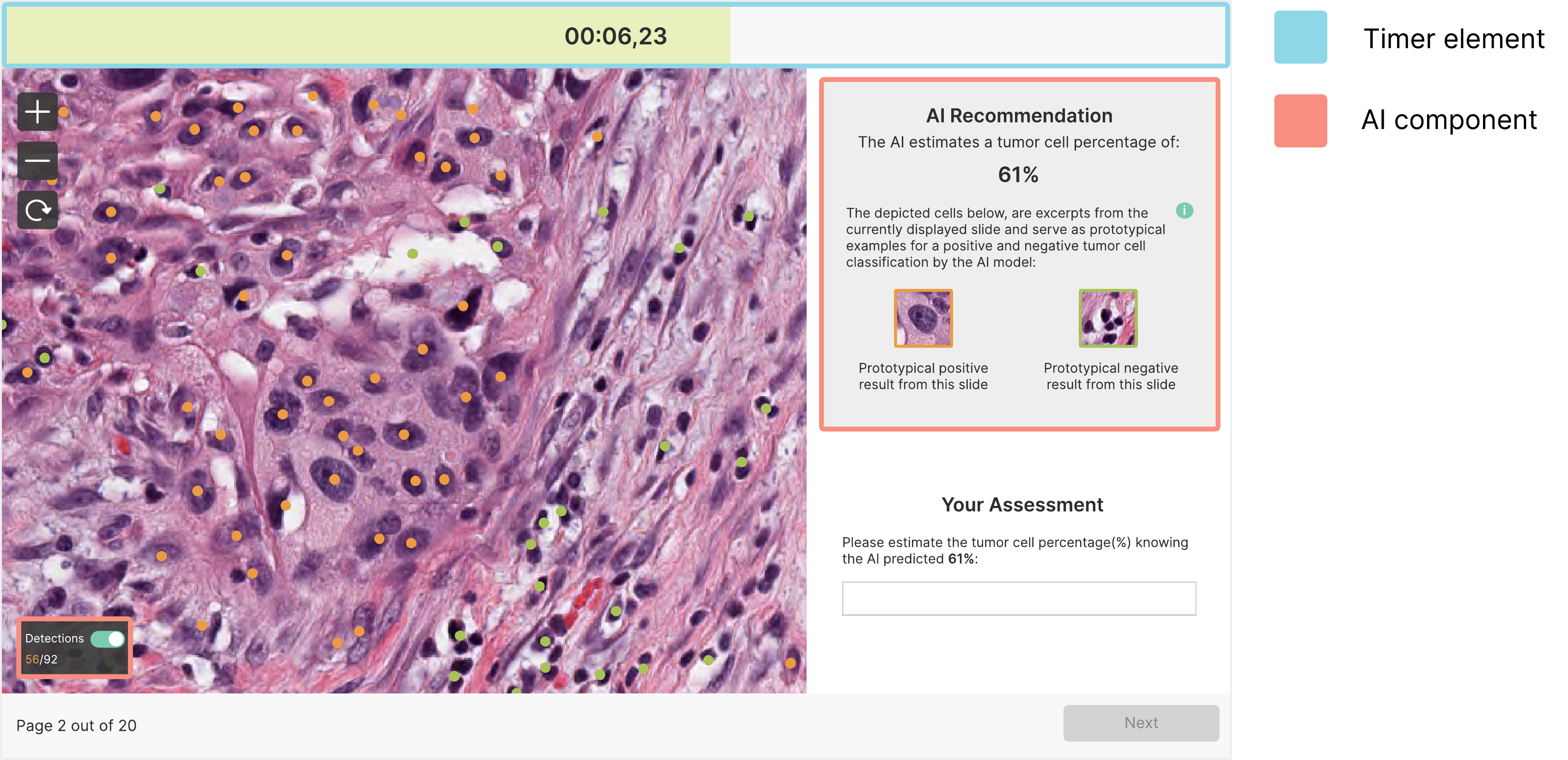}
    \caption{Study interface as seen by participants during TCP assessment. Depending on the experimental condition, the AI component and the reverse countdown timer element were made visible.}
    \label{fig:finalUI}
\end{figure}

\subsection{Experimental Design} \label{expDesign}
Based on the literature review the following hypotheses were established:

\noindent\textbf{H1:} The introduction of AI support, compared to the baseline (no AI), will lead to the emergence of new errors in the form of measurable AB (negative consultations).

\noindent\textbf{H2:} The presence of time constraints will increase both the frequency and severity of AB, as delineated in H1.

\noindent The study employed a $2\times2$ factorial, within-subject design with two independent variables (IVs): inclusion of AI (yes/no) and presence of time pressure (yes/no).

\textit{Automation bias.} The occurrence of AB constitutes the key dependent variable (DV). For this, instances of negative consultations, where an initially correct assessment in the baseline treatment ($\mathrm{Est}_\textrm{B}$) was altered to a false evaluation in the second round ($\mathrm{Est}_\textrm{AI}$) after exposure to an incorrect AI prediction ($\mathrm{Pred}_\textrm{AI}$), were quantified. To classify the correctness of TCP estimates (ranging from 0-100\%), we applied the 25\% threshold required for direct DNA sequencing \cite{Smits2014}. Assessments were deemed correct if they aligned with the ground truth (GT) in relation to this threshold. For the AB analysis, only dataset entries, where the independent estimate matches the GT, while both the AI recommendation and subsequent AI-aided assessment are erroneous, were considered. After filtering, the remaining data tuples were compared to the total number of AI-assisted TCP assessments to derive the AB occurrence rate.
% For the AB analysis, only dataset entries, containing the baseline, AI-assisted, and AI-derived TCP estimates for each slide per participant, where the independent estimate aligns with the GT, while both the AI recommendation and subsequent AI-aided assessment are erroneous, were considered. After filtering, the remaining data tuples were counted and compared to the total number of entries to calculate the AB occurrence rate.

\textit{Performance.}
Participant performance, calculated as the mean absolute deviation of TCP estimates from the GT ($\overline{\left |\mathrm{Dev}_\textrm{GT}\right |}$) per condition, is the second dependent variable.

\textit{Alignment with AI advice.} The relative dependence on AI recommendations was also measured using the judge advisor system \cite{sniezek1995cueing}, calculated per condition as $\overline{JAS} = \frac{1}{N_{\text{slides}}} \sum_{i=1}^{N_{\text{slides}}} \frac{\left | \mathrm{Est}_{\textrm{AI(i)}} - \mathrm{Est}_{\textrm{B(i)}} \right |}{\left | \mathrm{Est}_{\textrm{AI(i)}} - \mathrm{Pred}_{\textrm{AI(i)}} \right | + \left | \mathrm{Est}_{\textrm{AI(i)}} - \mathrm{Est}_{\textrm{B(i)}} \right |}$. The dependent variables, performance and AI-alignment, were analyzed for the entire dataset and the AB-filtered subset.

\textit{Procedure.} After a training session to acquaint participants with the user interface and study task, the experiment unfolded as follows: subjects were asked to estimate the TCP across 20 H\&E-stained slides, first independently and then with the aid of an AI. Identical image material was employed in both phases, separated by a two-week wash-out period to counteract learning effects. Half of the slides in each segment were subject to time stress, in the form of an expiring 10 second countdown. Since the time pressure (TP) conditions (no TP/with TP) were implemented consecutively in each segment, the study material was split into two image sets, consisting of 10 different yet similar slides in terms of TCP content and AI prediction quality. A within-group approach was adopted to maximize statistical power, given the challenge of recruiting experts. Subjects were randomly assigned to one of four groups, determining which TP condition and image set they begin both rounds with. Additionally, a balanced 10x10 Latin Square was employed to randomize the sequence of tissue slides within each image set.
 
\section{Results}
A total of 31 medical experts participated in the first round, with 28 advancing to the second phase, resulting in a final sample size of 28 subjects.

\textit{The general impact of AI integration and time pressure.}\footnote[2]{The normal distributions of all samples evaluated in this section were confirmed via Shapiro-Wilk tests.} A two-way repeated measures ANOVA conducted on the mean absolute deviation from the ground truth revealed a statistically significant main effect of AI integration on performance (F(1, 108) = 8.32, p = <.01, \(\alpha\) = 0.05). However, neither the main effect of time pressure (p = 0.19) nor the interaction between AI integration and TP (p = 0.46) were statistically significant. Examining the baseline in isolation, a one-tailed paired-samples t-test revealed a statistically significant decrease in performance under time stress (no TP: M = 13.55, SD = 4.81; with TP: M = 15.29, SD = 5.00; t(27) = -1.76, p = 0.045, \(\alpha\) = 0.05). Comparison of performance in the AI-aided treatments, albeit not statistically significant, reflects a similar trend (no TP: M = 11.74; with TP: M = 12.23). These results suggest that AI integration contributes to an overall performance increase, evidenced by a reduced $\overline{\left |\mathrm{Dev}_\textrm{GT}\right |}$, whereas TP seems to have the opposite effect. To evaluate the next IV, a two-tailed paired-samples t-test was conducted on the $\overline{JAS}$, revealing a statistically significant difference between the AI treatments with and without time constraints (no TP: M = 0.49, SD = 0.13; with TP: M = 0.55, SD = 0.12; t(27) = -2.80, p = <.01, \(\alpha\) = 0.05). This highlights that reliance on AI advice may intensify under time stress.

\textit{Automation bias.}
Due to reduced sample size after filtering (see section \ref{expDesign}), DVs did not consistently meet normality assumptions, hence the AB analysis will be conducted using descriptive statistics. Of 560 TCP estimates made with AI aid, practitioners adopted AI recommendations contradictory to their independent assessments, pertaining to the TCP being below/above the 25\% threshold, in only 67 cases. This included 29 positive consultations, where AI integration lead to correction of a previously erroneous decision, and 38 negative consultations, where an initially correct evaluation was overturned by incorrect AI guidance. The resulting AB rate consequently yielded approximately 7\% (38/560). Regarding the influence of TP, a comparison of the descriptive statistics in Table \ref{tab:descStats} indicates that the AB occurrence rate remains steady at approximately 7\%, irrespective of the presence of time constraints. However, time pressure appears to notably exacerbate the severity of AB, as evidenced by a decline in participant performance reflected in a heightened $\overline{\left |\mathrm{Dev}_\textrm{GT}\right |}$ (no TP: M = 19.42; with TP: M = 27.79). Additionally, another factor that might suggest an increased magnitude of AB under time stress is the greater alignment with AI advice, reflected in the rising $\overline{JAS}$ (no TP: M = 0.58; with TP: M = 0.65), when adopting the system's negative consultations under time strain.
\begin{table}[h]
   \caption{Descriptive statistics comparing automation bias occurrence (no. negative consultations), performance ($\overline{\left |Dev_\textrm{GT}\right |}$), and dependence on AI ($\overline{JAS}$) with and without time pressure.}
    \label{tab:descStats}
    \begin{tabular}{p{0.28\linewidth} p{0.30\linewidth} p{0.19\linewidth} p{0.19\linewidth}}
    \hline
      & \textbf{All negative consultations} & \textbf{No TP} & \textbf{With TP }\\
     Frequency  & 38/560 $\approx$ 7\% & 19/280 $\approx$ 7\% & 19/280 $\approx$ 7\%\\
     \hline
     & M $\mp$ SD & M $\mp$ SD & M $\mp$ SD\\
     \hline
     \rule{0pt}{3ex} Performance ($\overline{\left |Dev_\textrm{GT}\right |}$) & 23.61 $\mp$ 19.32 & 19.42 $\mp$ 17.37 & 27.79 $\mp$ 20.71\\
     $\overline{JAS}$ & 0.62 $\mp$ 0.29 & 0.58 $\mp$ 0.32 & 0.65 $\mp$ 0.25 \\
    \end{tabular}
    \end{table}
\section{Discussion and Conclusion}
Overall, integration of AI was found to improve participant performance. This ameliorating effect may also explain why the notable performance decline observed under time pressure in the baseline condition was less pronounced during AI-aided evaluations. Additionally, TP appeared to increase alignment with AI advice, likely due to the strain on cognitive resources under stress, which could be considered beneficial when the AI is accurate, but detrimental when the system errs. Interestingly, despite AI’s potential, we recorded that pathologists were largely unwilling to adopt model recommendations, that contradicted their initial judgments, regardless of the correctness of the system's output. This suggests that AB may not be the primary cognitive bias in AI-aided medical decision-making. Nonetheless, by quantifying the number of accepted negative consultations, we determined an AB occurrence rate of approximately 7\% in its "purest" form. This aligns with existing empirical research on AB \cite{Goddard2012}, reporting negative consultation acceptance rates of 6\% to 11\%. As our findings suggest that interaction with AI can indeed induce AB, leading to errors, that would not have occurred in the absence of system guidance, hypothesis 1 is fully accepted. Contrary to the expectations outlined in H2, the introduction of time pressure did not affect the occurrence frequency of AB. This could be attributed to our modest sample of AB incidents, which might not have been sufficient to capture the effect in question, as well as the nature of our TP simulation. In practice, deadlines manifest in form of volumes of specimen to be assessed rather than individual countdowns, in turn, participants' reactions may not fully reflect their responses under real clinical time constraints. Consistent with our general observations, time pressure appeared to increase dependence on negative system consultations. This heightened alignment with erroneous AI output might have exacerbated the severity of AB, mirrored in the pronounced performance decline, beyond what would be expected from the negative influence of TP on performance alone, as seen in the general analysis. In summary, while we did not observe significant effects of time strain on AB occurrence, our results indicate that its severity worsens under time stress. Thus, we partially accept hypothesis 2. It has to be acknowledged, that due to the limited availability of expert participants, the study was conducted on a modest sample (n=28). Although a within-subject design was adopted to augment statistical power, the effects showcased may be under-/over-represented due to sample variability compared to the target population. Moreover, interface design choices such as the omission of clinical background information may have diminished task realism and altered participant behavior, prompting them to approach the study with less diligence as their everyday examinations, potentially impacting observable AB rates. Despite this, the findings serve as a valuable step towards a more holistic understanding of AB in AI-aided medical decision-making and its influencing factors. By highlighting the risk of cognitive biases, we aim to support the safe integration of AI in critical fields like healthcare. Future work could explore the effectiveness of debiasing strategies in minimizing AI-induced cognitive biases such as AB, specifically evaluating their utility under time stress.

\begin{acknowledgement}
This work was supported by the Bavarian Institute for Digital Transformation (bidt) through the “ReGInA” grant.
\end{acknowledgement}

\printbibliography
\end{document}